\documentclass[11pt]{article}
\usepackage{exscale}
\usepackage[final]{epsfig}
\usepackage{epsf}
\usepackage{picins}
\usepackage{amssymb,amsmath}

\begin{document}

\title{A Generalized Sampling Theorem for Frequency Localized Signals}

\author{Edwin Hammerich\\
    \small Ministry of Defence\\
    \small Kulmbacher Str. 58-60, D-95030 Hof, Germany\\
    \small e-mail: edwin.hammerich@ieee.org}
\date{}
\maketitle
\begin{center}Dedicated to Professor Abdul J. Jerri on the occasion of his 75th birthday\end{center}
\begin{abstract}
A generalized sampling theorem for frequency localized signals is presented. The generalization in the
proposed model of sampling is twofold: (1) It applies to various prefilters effecting a "soft"
bandlimitation, (2) an approximate reconstruction from sample values rather than a perfect one is 
obtained (though the former might be "practically perfect" in many cases). For an arbitrary finite-energy
signal the frequency localization is performed by a prefilter realizing a crosscorrelation with a function 
of prescribed properties. The range of the prefilter, the so-called localization space, is described in 
some detail. Regular sampling is applied and a reconstruction formula is given. For the reconstruction 
error a general error estimate is derived and connections between a critical sampling interval 
and notions of "soft bandwidth" for the prefilter are indicated. Examples based on the sinc-function, 
Gaussian functions and B-splines are discussed.
\end{abstract}

\newtheorem{definition}{Definition}
\newtheorem{lemma}{Lemma}
\newtheorem{proposition}{Proposition}
\newtheorem{theorem}{Theorem}
\newtheorem{corollary}{Corollary}
\newtheorem{remark}{Remark}

\section{Introduction}
In time-frequency analysis the concept of time-frequency localization is well-known \cite{Groch}, 
\cite{Daubechies}, \cite{Fei}. It means the approximate concentration of a signal in both time and 
frequency allowing for tails of strong decay in either dimension. In many areas of communications the 
concept of (strict) bandlimitation is still prevailing. One reason might be the classical sampling 
theorem of Whittaker, Kotel'nikov and Shannon \cite{Shannon}, \cite{Jerri1}, \cite{Butzer1}, 
\cite{Unser}, \cite{Vaid} where bandlimited signals are the basic assumption. This leads to difficulties 
in practical applications \cite{Unser} of the classical sampling theorem and researchers have been 
trying to overcome the entailed complications \cite{Blu99a}, \cite{Butzer2}, \cite{Eldar}, 
\cite{Jerri2}, \cite{Sun}, \cite{UnserAldroubi2}, \cite{Unser}. The goal 
of the present work is to contribute to these attempts by loosening the assumption of strict 
bandlimitation while 
retaining the model of sampling of the classical sampling theorem: An arbitrary finite-energy input 
signal is preprocessed by a prefilter, the filter output signal lying in a well-structured space similar 
to the space of bandlimited signals is ideally sampled at equidistant points of time, finally the 
complete filter output signal is to be reconstructed. A whole variety of prefilters is described 
some of which are presumably suitable for practical applications. The price to be paid will be, in 
general, imperfect reconstruction. Some effort is made to estimate the incurred reconstruction error and 
to find criteria for the size of the sampling interval to guarantee good reconstruction. This is well in
the sense of A. J. Jerri's work on error analysis in sampling theory 
and applications \cite{Jerri1}, \cite{Jerri2}. The chosen approach incorporates from the start some 
measure of generalized bandwidth for the used prefilter so that eventually the link between that 
generalized bandwidth (the so-called "soft bandwidth") and a critical sampling interval shall be found. 
That critical sampling interval would then correspond to the Nyquist interval in case of the classical 
sampling theorem \cite{Shannon}, \cite{Vaid}.

The paper is organized as follows. In Section 2 the frequency localization operator (prefilter) is 
defined and basic assumptions are compiled. In Section 3 the so-called localization space is described. 
In Section 4 a perfect reconstruction formula for elements of some subspace of localization space is 
derived. In Section 5 a general error estimate for the reconstruction error is given. As a conclusion, 
in Section 6 the sampling theorem is presented. In subsections at the end of Section 5 and 6 examples 
are discussed.

The following notation is used: $L^2(\mathbb{R})$ is the space of square integrable functions 
(or finite-energy signals) $f:\mathbb{R}\rightarrow\mathbb{C}$ with inner product 
$\langle f_1,f_2\rangle=\int_{-\infty}^\infty f_1(x)\overline{f_2(x)}\,dx.$ For the Fourier transform we 
adopt the convention $\hat{f}(\xi)=(2\pi)^{-1/2}\int_{-\infty}^\infty e^{-ix\xi}f(x)\,dx$, where $x$
denotes time and $\xi$ (angular) frequency. $\mathbb{R}_+$ denotes the set of positive real numbers.

\section{Frequency Localization Operator}
For an arbitrary finite-energy signal $f\in L^2(\mathbb{R})$ frequency localization is performed by 
an operator $\boldsymbol{P}_\varphi:L^2({\mathbb R})\rightarrow L^2({\mathbb R})$ given by
\begin{equation}
   (\boldsymbol{P}_\varphi f)(x)=\int_{-\infty}^\infty f(y)\overline{\varphi(y-x)}\,dy
   =\int_{-\infty}^\infty e^{ix\xi}\,\overline{\hat{\varphi}(\xi)}\hat{f}(\xi)\,d\xi. \label{OpA}
\end{equation}
We shall call $\boldsymbol{P}_\varphi$ \textit{prefilter} because in practice it would be an anti-aliasing
prefilter or a related nonideal acquisition device \cite{UnserAldroubi2}. Since 
$(\boldsymbol{P}_\varphi f)(x)=\langle f,\varphi(\cdot-x)\rangle,\,x\in\mathbb{R}$, the 
prefilter realizes a crosscorrelation of the input signal $f$ with $\varphi$ (rather than a
convolution with $\overline{\varphi(-x)}$). We make the following assumptions on the 
\textit{prefilter function} $\varphi$:

(i) $\varphi\in L^2({\mathbb R})$

(ii) It holds the generalized moment condition 
\begin{equation}
   M_w(\varphi)=\int_{-\infty}^\infty w(|\xi|)|\hat{\varphi}(\xi)|^2\,d\xi<\infty,          \label{mom}
\end{equation}

where $w=w(\xi)$ for $\xi>0$  is a positive and monotonically increasing weight

function of sufficient growth,
\begin{equation}
   w(\xi)\ge c\xi^{1+\epsilon}\:\:\:\forall\xi\in[1,\infty)\:(c>0,\,\epsilon>0).    \label{growth}
\end{equation}

(iii) For any $\lambda\in\mathbf{\Lambda}$, where $\mathbf{\Lambda}$ is some non-empty subset of 
$\mathbb{R}_+$, the family

of functions $\{\varphi(\cdot-n\lambda);n\in\mathbb{Z}\}$ forms a Riesz basis 
in $L^2(\mathbb{R})$.

\noindent We refer to \cite{Christensen}, \cite{Unser} concerning Riesz bases and their relevance to 
signal processing. The Riesz basis condition is equivalent to the existence of positive real numbers 
$A,\,B$ (possibly depending on $\lambda$) so that   
\begin{equation}
  0<A\le \mathnormal{\Lambda}\sum_{n=-\infty}^\infty 
       |\hat{\varphi}(\xi+n\mathnormal{\Lambda})|^2\le B<\infty\:\:\:\mbox{a.e.}, \label{Riesz}
\end{equation}
where $\mathnormal{\Lambda}=2\pi/\lambda$. See \cite{UnserAldroubi1} for a proof in case $\lambda=1$,  
the argument carries over to arbitrary $\lambda>0$ without changes. Actually, we shall use the Riesz 
basis condition mostly in form of Ineq. (\ref{Riesz}). 

Because of the upper bound in Ineq. (\ref{Riesz}) the Fourier transform $\hat{\varphi}$ is bounded almost
everywhere on $\mathbb{R}$ so that the range of $\boldsymbol{P}_\varphi$ is a subset of 
$L^2(\mathbb{R})$ as presumed.
We note that because of condition (\ref{growth}) the prefilter function $\varphi$ is in a Sobolev space 
\cite{Mazja} $H^r(\mathbb{R})=\{u\in L^2(\mathbb{R});\int|\hat{u}(\xi)|^2(1+|\xi|^2)^r\,d\,\xi<\infty\}$ 
of order $r>\frac{1}{2}$. Thus, by reason of the Sobolev embedding theorem, $\varphi$ will 
necessarily be a continuous function.

In our paper monomial weights $w(\xi)=|\xi|^s,\,s>1$, and Gaussian weights $w(\xi)=\exp(s\xi^2),\,s>0$, 
will be used. As prefilter functions $\varphi$ the sinc-function, Gaussian functions and B-splines will 
be taken. 

\section{Localization Space}
The \textit{localization space} $\mathcal{P}_\varphi$ of the prefilter 
$\boldsymbol{P}_\varphi$ corresponds to the space of bandlimited signals in case of an ideal low-pass
filter and is defined as the range of $\boldsymbol{P}_\varphi$,
\[
   \mathcal{P}_\varphi=\{g=\boldsymbol{P}_\varphi f;\,f\in L^2({\mathbb R})\}.
\]
$\mathcal{P}_\varphi$ is a linear, not necessarily closed subspace of $L^2({\mathbb R})$, it even may be 
dense in $L^2({\mathbb R})$. Moreover, the localization space is invariant with respect to arbitrary 
translations $g\mapsto g(\cdot-t),\,t\in\mathbb{R}$. The function $\mathnormal{\Phi}\in \mathcal{P}_\varphi$ 
defined by
\[
   \mathnormal{\Phi}=\boldsymbol{P}_\varphi\varphi\stackrel{\mathrm{Fourier}}{\longleftrightarrow}
                                          \hat{\mathnormal{\Phi}}(\xi)=\sqrt{2\pi}|\hat{\varphi}(\xi)|^2
\]
will be of importance. Since 
$\mathnormal{\Phi}(x)=\langle \varphi,\varphi(\cdot-x)\rangle,\,x\in\mathbb{R},$ it is the 
autocorrelation function of $\varphi$. On $\mathcal{P}_\varphi$ we define the inner product
\begin{equation}
   \langle g_1,g_2\rangle_\varphi=\frac{1}{2\pi}
                \int_{\mathrm{supp}\,\hat{\varphi}} \hat{g}_1(\xi)\overline{\hat{g}_2(\xi)}\,
		              |\hat{\varphi}(\xi)|^{-2}d\xi  \label{ip}
\end{equation}
with corresponding norm $\|g\|_\varphi=\sqrt{\langle g,g \rangle_\varphi}$.

The reproducing kernel Hilbert space (RKHS) property of localization space $\mathcal{P}_\varphi$ 
will be described in the next theorem. The shorthand notation 
$L^2({\mathbb R}|\,\mathrm{supp}\, \hat{\varphi})$ for the closed subspace
$\{f\in L^2({\mathbb R});\,\mathrm{supp}\,\hat{f}\subseteq \mathrm{supp}\,\hat{\varphi}\}$ of 
$L^2({\mathbb R})$ will be used. We remark that 
$L^2({\mathbb R}|\,\mathrm{supp}\, \hat{\varphi})=L^2({\mathbb R})$ in case of a Gaussian function or a 
B-spline $\varphi$ (see Section 6.1). 
\begin{proposition} \label{Propo_RKHS} The linear space $\mathcal{P}_\varphi$ endowed with the inner
product (\ref{ip}) is an RKHS. The reproducing kernel is 
$K(x,y)=\mathnormal{\Phi}(y-x)$. $\boldsymbol{P}_\varphi$ maps 
$L^2({\mathbb R}|\,\mathrm{supp}\,\hat{\varphi})$ isometrically onto $\mathcal{P}_\varphi$, i.e., if 
$g_i=\boldsymbol{P}_\varphi f_i$ with $f_i\in L^2({\mathbb R}|\,\mathrm{supp}\,\hat{\varphi}),i=1,2,$ 
then 
\begin{equation}
   \langle g_1,g_2\rangle_\varphi =\langle f_1,f_2\rangle.          \label{isom}
\end{equation}
\end{proposition}
\textit{Proof.} Since the Fourier transform of $g_i=\boldsymbol{P}_\varphi f_i$ is 
$\hat{g}_i(\xi)=\sqrt{2\pi}\,\overline{\hat{\varphi}(\xi)}\hat{f}_i(\xi),\,\xi\in\mathbb{R}$, 
we have by Parseval's formula that 
$\langle g_1,g_2 \rangle_\varphi=\langle\hat{f}_1,\hat{f}_2\rangle=\langle f_1,f_2 \rangle$. 
If $g=\boldsymbol{P}_\varphi f$ with $f\in L^2(\mathbb{R})$, then the function $f_0$ defined by 
$\hat{f}_0(\xi)=\hat{g}(\xi)/(\sqrt{2\pi}\,\overline{\hat{\varphi}(\xi)})$ in case 
$\xi\in\mathrm{supp}\,\hat{\varphi}$ and $\hat{f}_0(\xi)=0$ else is in 
$L^2(\mathbb{R}|\,\mathrm{supp}\,\hat{\varphi})$ and $g=\boldsymbol{P}_\varphi f_0$. Thus, 
$\boldsymbol{P}_\varphi L^2(\mathbb{R}|\,\mathrm{supp}\,\hat{\varphi})=\mathcal{P}_\varphi$. 
Because of the isometry relation (\ref{isom}), $\mathcal{P}_\varphi$ endowed with the inner product 
$\langle \cdot,\cdot \rangle_\varphi $ becomes a Hilbert space as is 
$L^2(\mathbb{R}|\,\mathrm{supp}\,\hat{\varphi})$ with respect to the inner product 
$\langle \cdot,\cdot \rangle$ of $L^2(\mathbb{R})$.

For any 
$x\in\mathbb{R}$ we define the function $\varphi_x\in L^2(\mathbb{R}|\,\mathrm{supp}\,\hat{\varphi})$ by 
$\varphi_x(y)=\varphi(y-x),\,y\in\mathbb{R}$. Since 
$(\boldsymbol{P}_\varphi\varphi_x)(y)=\mathnormal{\Phi}(y-x)=K(x,y)$, we have 
$K(x,\cdot)\in \mathcal{P}_\varphi$. For any function $g=\boldsymbol{P}_\varphi f_0$ with 
$f_0\in L^2(\mathbb{R}|\,\mathrm{supp}\,\hat{\varphi})$ and any $x\in\mathbb{R}$ we now infer by means 
of (\ref{isom}) that
\begin{eqnarray*}
	\langle g,K(x,\cdot)\rangle_\varphi &=&\langle \boldsymbol{P}_\varphi f_0,
	                                            \boldsymbol{P}_\varphi\varphi_x\rangle_\varphi \\
				&=&\langle f_0,\varphi_x\rangle\\
				&=&\int_{-\infty}^\infty f_0(y)\overline{\varphi(y-x)}\,dy\\
				&=&(\boldsymbol{P}_\varphi f_0)(x)\\
				&=&g(x).
\end{eqnarray*}
Thus, $K(x,y)$ is the reproducing kernel in the Hilbert space $\mathcal{P}_\varphi$. This concludes 
the proof of Proposition \ref{Propo_RKHS}.

We remark that Eq. (\ref{isom}) should not be taken as a
definition of the inner product $\langle g_1,g_2\rangle_\varphi$, because it implies inversion of 
operator $\boldsymbol{P}_\varphi$ which, in general, is an ill-posed problem. We shall need the following 
subspaces of domain $L^2(\mathbb{R})$ and range $\mathcal{P}_\varphi$ of operator 
$\boldsymbol{P}_\varphi$.

\begin{definition} The shift-invariant subspace $V_\lambda(\varphi)\subseteq L^2(\mathbb{R})$ is the 
closed linear span of $\{\varphi_n=\varphi(\cdot-n\lambda);\,n\in{\mathbb Z}\}$ in $L^2(\mathbb{R})$.
\end{definition}
\begin{definition} The subspace $\mathcal{R}_\lambda(\varphi)\subseteq\mathcal{P}_\varphi$ 
("reconstruction subspace") is the closed linear span of 
$\{\mathnormal{\Phi}_n=\mathnormal{\Phi}(\cdot-n\lambda);\,n\in{\mathbb Z}\}$ in $\mathcal{P}_\varphi$.
\end{definition}
\noindent\textbf{Remark 1} The shift-invariant spaces $V_T(\varphi),\,T>0,$ of Unser \cite{Unser} are
generated by translated \textit{and dilated} versions 
$\varphi\left(\frac{x}{T}-n\right),\,n\in\mathbb{Z},$ of a generating function $\varphi$.
\begin{proposition} \label{prop_struct} $\boldsymbol{P}_\varphi$ is an isometry from 
$V_\lambda(\varphi)$ onto $\mathcal{R}_\lambda(\varphi)$. If $\{\mathnormal{\Phi}_n;n\in\mathbb{Z}\}$ 
forms a Riesz basis in $L^2(\mathbb{R})$, then $\mathcal{R}_\lambda(\varphi)$ may be identified 
with the shift-invariant subspace $V_\lambda(\mathnormal{\Phi})\subseteq L^2(\mathbb{R})$ defined as the 
closed linear span of $\{\mathnormal{\Phi}_n;n\in\mathbb{Z}\}$ in $L^2(\mathbb{R})$.
\end{proposition}
\textit{Proof.} 
Since $\boldsymbol{P}_\varphi \varphi_n=\mathnormal{\Phi}_n\,\forall n\in\mathbb{Z}$ and, by 
Proposition \ref{Propo_RKHS}, $\boldsymbol{P}_\varphi$ is an isometry from 
$V_\lambda(\varphi)\subseteq L^2(\mathbb{R}|\,\mathrm{supp}\,\hat{\varphi})$ into $\mathcal{P}_\varphi$, 
it follows that $\boldsymbol{P}_\varphi V_\lambda(\varphi)=\mathcal{R}_\lambda(\varphi)$.

An element $v(x)=\sum_{n\in\mathbb{Z}}^\mathrm{finite}c_n\mathnormal{\Phi}_n$ of the linear 
span of $\{\mathnormal{\Phi}_n;n\in\mathbb{Z}\}$ has norms in $\mathcal{P}_\varphi$ and 
$L^2(\mathbb{R})$, resp., 
\[
   \|v\|_\varphi=\left[\int_0^1|C(\xi)|^2A_\varphi(\mathnormal{\Lambda}\xi)\,d\xi\right]^{1/2},\,
   \|v\|=\left[\int_0^1|C(\xi)|^2A_\mathnormal{\Phi}(\mathnormal{\Lambda}\xi)\,d\xi\right]^{1/2},
\]
where $\mathnormal{\Lambda}=2\pi/\lambda,\,C(\xi)=\sum_{n\in\mathbb{Z}}^\mathrm{finite}c_ne^{-2\pi in\xi}$, 
$A_\varphi(\xi)$ is the central term in (\ref{Riesz}), and  
$A_\mathnormal{\Phi}(\xi)=\mathnormal{\Lambda}\sum_{k\in\mathbb{Z}}|\hat{\mathnormal{\Phi}}
(\xi+k\mathnormal{\Lambda})|^2$. Because of inequality (\ref{Riesz}) we have norm equivalence 
$\|v\|_\varphi\asymp [\int_0^1|C(\xi)|^2\,d\xi]^{1/2}$. If $\{\mathnormal{\Phi}_n;n\in\mathbb{Z}\}$ also 
forms a Riesz basis, a similar inequality will hold for $A_\mathnormal{\Phi}(\xi)$, so that 
$\|v\|\asymp[\int_0^1|C(\xi)|^2\,d\xi]^{1/2}$. Then, $\|v\|_\varphi\asymp\|v\|$ and consequently 
$\mathcal{R}_\lambda(\varphi)=V_\lambda(\mathnormal{\Phi})$, which concludes the proof of Proposition
\ref{prop_struct}.

In case of a Gaussian function $\varphi$, $\{\mathnormal{\Phi}_n;n\in\mathbb{Z}\}$ indeed forms a Riesz
basis in $L^2(\mathbb{R})$ \cite{Hammerich2}. But examples can be given where, albeit $\varphi$ 
satisfies all the assumptions made in Section 2, we have
\[ 
   0<A_\mathnormal{\Phi}(\xi)\le \frac{1}{k}\:\:\:\mbox{a.e.}\:\:\:\xi \in S_k
\]
for some $\lambda\in\mathbf{\Lambda}$ and a sequence of subsets 
$S_k\subseteq \mathbb{R},\,k=1,2,\ldots$, of positive measure. As a consequence, 
$\{\mathnormal{\Phi}_n=\mathnormal{\Phi}(\cdot-n\lambda);n\in\mathbb{Z}\}$ neither can form a Riesz 
basis nor a frame in $L^2(\mathbb{R})$, cf. \cite[Theorem 7.2.3]{Christensen}.
\begin{proposition} \label{Decomp} Localization space $\mathcal{P}_\varphi$ has the orthogonal 
decomposition 
\begin{equation}
  \mathcal{P}_\varphi=\mathcal{R}_\lambda(\varphi) \oplus \mathcal{N}_\lambda(\varphi),\label{decomp}
\end{equation} 
where subspace $\mathcal{N}_\lambda(\varphi)$ consists of all functions 
$h\in \mathcal{P}_\varphi$ with the property that $h(n\lambda)=0\:\forall n\in\mathbb{Z}$. Every function 
$g\in \mathcal{R}_\lambda(\varphi)$ is completely determined by the sample values 
$g(n\lambda),\,n\in\mathbb{Z}$.
\end{proposition}
\textit{Proof.} The space $L^2(\mathbb{R})$ has the orthogonal decomposition
\begin{equation}
   L^2(\mathbb{R})=V_\lambda(\varphi)\oplus V_\lambda(\varphi)^\perp.           \label{deco1}
\end{equation}
Application of $\boldsymbol{P}_\varphi$ yields 
$\mathcal{P}_\varphi=\boldsymbol{P}_\varphi V_\lambda(\varphi)+
\boldsymbol{P}_\varphi(V_\lambda(\varphi)^\perp)$. 
By Proposition \ref{prop_struct}, 
$\boldsymbol{P}_\varphi V_\lambda(\varphi)=\mathcal{R}_\lambda(\varphi)$. We now show that 
$\boldsymbol{P}_\varphi(V_\lambda(\varphi)^\perp)=\mathcal{N}_\lambda(\varphi)$. If 
$f\in V_\lambda(\varphi)^\perp$, then $h=\boldsymbol{P}_\varphi f$ satisfies
\begin{equation}
  h(n\lambda)=(\boldsymbol{P}_\varphi f)(n\lambda)
              =\langle f,\varphi(\cdot-n\lambda)\rangle=0\:\:\:\forall n\in\mathbb{Z},  \label{null}
\end{equation}
So, $h\in \mathcal{N}_\lambda(\varphi)$. Conversely, if $h\in \mathcal{N}_\lambda(\varphi)$, then 
$h=\boldsymbol{P}_\varphi f$ for some $f\in L^2(\mathbb{R})$. Reading (\ref{null}) from the left to the
right shows that $f$ is orthogonal on the dense subset $\{\varphi_n;n\in\mathbb{Z}\}$ of 
$V_\lambda(\varphi)$. So, $f\in V_\lambda(\varphi)^\perp$. 

The orthogonality $\mathcal{R}_\lambda(\varphi)\perp\mathcal{N}_\lambda(\varphi)$ in 
$\mathcal{P}_\varphi$ (implying $\mathcal{R}_\lambda(\varphi)\cap \mathcal{N}_\lambda(\varphi)=\{0\}$) 
is seen as follows. If $g\in\mathcal{R}_\lambda(\varphi),\,h\in\mathcal{N}_\lambda(\varphi)$, then 
$g=\boldsymbol{P}_\varphi f_1$ for some 
$f_1\in V_\lambda(\varphi)\subseteq L^2(\mathbb{R}|\,\mathrm{supp}\,\hat{\varphi})$ and 
$h=\boldsymbol{P}_\varphi f_2$ for some $f_2\in V_\lambda(\varphi)^\perp$. Define 
$f_0\in L^2(\mathbb{R}|\,\mathrm{supp}\,\hat{\varphi})$ by 
$\hat{f}_0=\chi_{\mathrm{supp}\,\hat{\varphi}}\cdot \hat{f}_2$. Then 
$\boldsymbol{P}_\varphi f_2=\boldsymbol{P}_\varphi f_0$, So, by Proposition \ref{Propo_RKHS}, we have 
$\langle g,h \rangle_\varphi
=\langle \boldsymbol{P}_\varphi f_1,\boldsymbol{P}_\varphi f_0 \rangle_\varphi
=\langle f_1,f_0 \rangle=\langle \hat{f}_1,\hat{f}_0 \rangle=\langle \hat{f}_1,\hat{f}_2 \rangle
=\langle f_1,f_2 \rangle=0$. Thus, $g\perp h$ in $\mathcal{P}_\varphi$. 

Finally, if $g_1,\,g_2\in \mathcal{R}_\lambda(\varphi)$ with 
$g_1(n\lambda)=g_2(n\lambda)\,\forall n\in\mathbb{Z}$, then 
$g_1-g_2\in \mathcal{R}_\lambda(\varphi)\cap \mathcal{N}_\lambda(\varphi)=\{0\}$. Hence, $g_1=g_2$, 
which concludes the proof of Proposition \ref{Decomp}.

We refer to \cite{Eldar}, \cite{EldarWerther}, \cite{EldarDvork} where orthogonal decompositons of
Hilbert spaces have been used for the purpose of sampling and reconstruction in a more general and 
abstract setting. 

\section{Perfect Reconstruction in a Subspace}
Since $\{\mathnormal{\Phi}_n;n\in\mathbb{Z}\}$ not always forms a Riesz basis, the proof of the 
next theorem is based on frame theory. Concerning frame theory we refer to \cite{Christensen}, 
\cite{Groch}, \cite{Young}. Because $\{\mathnormal{\Phi}_n;n\in\mathbb{Z}\}$ also not always forms
a frame for its closed linear span \textit{in} $L^2(\mathbb{R})$ we cannot resort to, e.g., 
\cite[Theorem 2.1]{Sun}. The RKHS property of $\mathcal{P}_\varphi$ will now prove helpful; see  
\cite{Kramer}, \cite{NasWalt}, \cite{Walter} with regard to application of reproducing kernel Hilbert
spaces in sampling theory. 
\begin{theorem} \label{Theo_perfrec} Let the prefilter function $\varphi$ and the set 
$\mathbf{\Lambda}\subseteq\mathbb{R}_+$ be as assumed in Section 2 and let 
$\lambda\in \mathbf{\Lambda}$. Then any function $g\in \mathcal{R}_\lambda(\varphi)$ can be perfectly 
reconstructed from its sample values $g(n\lambda),\,n\in \mathbb{Z},$ by the series
\begin{equation}
   g(x)=\sum_{n=-\infty}^\infty g(n\lambda)\mathnormal{\Phi}_\mathrm{int}(x-n\lambda),\,
                                                                       x\in\mathbb{R},   \label{series}
\end{equation}
where the interpolating function $\mathnormal{\Phi}_\mathrm{int}\in \mathcal{R}_\lambda(\varphi)$ is 
given by
\begin{equation}
   \hat{\mathnormal{\Phi}}_\mathrm{int}(\xi)=\frac{\hat{\mathnormal{\Phi}}(\xi)}{\frac{\mathnormal{\Lambda}}{\sqrt{2\pi}}
                                   \sum_{n\in {\mathbb Z}}\hat{\mathnormal{\Phi}}(\xi+n\mathnormal{\Lambda})},\, 
				           \mathnormal{\Lambda}=\frac{2\pi}{\lambda}.    \label{Phi_int}
\end{equation}
The series in (\ref{series}) converges in the norm of $\mathcal{P}_\varphi$ and uniformly on $\mathbb{R}$. 
\end{theorem}
\textit{Proof.} Since $\{\varphi_n;n\in \mathbb{Z}\}$ is a Riesz basis of 
$V_\lambda(\varphi)$ it also forms a frame for that space. Hence, there exist positive constants $A,\,B$ 
(actually the same as in (\ref{Riesz})) so that for all $f\in V_\lambda(\varphi)$ we have
\[
   A\,\|f\|^2\le \sum_{n\in\mathbb{Z}} |\langle f,\varphi_n \rangle|^2\le B\,\|f\|^2.
\]
Because $\mathnormal{\Phi}_n=\boldsymbol{P}_\varphi \varphi_n$ and 
$\boldsymbol{P}_\varphi:V_\lambda(\varphi)\rightarrow \mathcal{R}_\lambda(\varphi)$ is an isometry we 
conclude that for all $g\in \mathcal{R}_\lambda(\varphi)$ it holds that
\[
   A\,\|g\|_\varphi^2\le \sum_{n\in\mathbb{Z}} |\langle g,\mathnormal{\Phi}_n \rangle_\varphi|^2
                                                                        \le B\,\|g\|_\varphi^2.
\]
Thus, $\{\mathnormal{\Phi}_n;n\in \mathbb{Z}\}$ is a frame for the Hilbert space 
$\mathcal{R}_\lambda(\varphi)\subseteq \mathcal{P}_\varphi$. The corresponding frame 
operator $S:\mathcal{R}_\lambda(\varphi)\rightarrow \mathcal{R}_\lambda(\varphi)$ has the representation  
$Sg=\sum_{n\in\mathbb{Z}}\langle g,\mathnormal{\Phi}_n\rangle_\varphi \mathnormal{\Phi}_n$. By frame 
theory it has a continuous inverse 
$S^{-1}:\mathcal{R}_\lambda(\varphi)\rightarrow \mathcal{R}_\lambda(\varphi)$ yielding the 
reconstruction formula
\begin{equation}
   g=\sum_{n\in\mathbb{Z}}\langle g,\mathnormal{\Phi}_n\rangle_\varphi S^{-1}\mathnormal{\Phi}_n.
                                                                                    \label{frame_reco}
\end{equation}
Because of the RKHS property of the localization space $\mathcal{P}_\varphi$ we already know that 
$\langle g,\mathnormal{\Phi}_n\rangle_\varphi=
\langle g,\mathnormal{\Phi}(\cdot-n\lambda)\rangle_\varphi=g(n\lambda)$. The uniquely determined 
solution $\varphi_\mathrm{dual}\in V_\lambda(\varphi)$ of the system of equations
\[
   \langle \varphi_\mathrm{dual},\varphi(\cdot-n\lambda)\rangle=\delta_n,\,n\in\mathbb{Z},
\]
exists in virtue of the positive lower bound in Ineq. (\ref{Riesz}) and has Fourier transform
\begin{equation}
   \hat{\varphi}_\mathrm{dual}(\xi)=\frac{\hat{\varphi}(\xi)}{\mathnormal{\Lambda}\sum_{k\in{\mathbb Z}}
                                            |\hat{\varphi}(\xi+k\mathnormal{\Lambda})|^2}. \label{dualf2}
\end{equation}
(The argument in \cite{Hammerich2} in case of a Gaussian function $\varphi$ carries over to the general 
case without changes.) Defining 
$\mathnormal{\Phi}_\mathrm{int}=\boldsymbol{P}_\varphi \varphi_\mathrm{dual}$ we infer, again by 
isometry, that
\begin{eqnarray*}
   S[\mathnormal{\Phi}_\mathrm{int}(\cdot-m\lambda)]
     &=&\sum_{n\in\mathbb{Z}}\langle \mathnormal{\Phi}_\mathrm{int}(\cdot-m\lambda),
                                                \mathnormal{\Phi}_n\rangle_\varphi \mathnormal{\Phi}_n\\
     &=&\sum_{n\in\mathbb{Z}}\langle \boldsymbol{P}_\varphi(\varphi_\mathrm{dual}(\cdot-m\lambda)),
                            \boldsymbol{P}_\varphi\varphi_n\rangle_\varphi \mathnormal{\Phi}_n\\
     &=&\sum_{n\in\mathbb{Z}}\langle \varphi_\mathrm{dual}(\cdot-m\lambda),
                                                    \varphi(\cdot-n\lambda)\rangle \mathnormal{\Phi}_n\\
     &=&\sum_{n\in\mathbb{Z}} \delta_{mn} \mathnormal{\Phi}_n\\
     &=&\mathnormal{\Phi}_m.
\end{eqnarray*}
Thus, $S^{-1}\mathnormal{\Phi}_m=\mathnormal{\Phi}_\mathrm{int}(\cdot-m\lambda)$ and 
(\ref{frame_reco}) turns into (\ref{series}). By means of the Fourier domain representation of 
$\boldsymbol{P}_\varphi$ we readily obtain
\[
   \mathnormal{\Phi}_\mathrm{int}(x)=\frac{1}{\sqrt{2\pi}}\int_{-\infty}^\infty e^{ix\xi}
               \frac{|\hat{\varphi}(\xi)|^2}{\frac{\mathnormal{\Lambda}}{\sqrt{2\pi}}
                      \sum_{n\in {\mathbb Z}}|\hat{\varphi}(\xi+n\mathnormal{\Lambda})|^2}\,d\xi,
\]
which proves (\ref{Phi_int}).

Frame theory ensures convergence of the series in (\ref{frame_reco}) in the norm 
$\|\cdot\|_\varphi$. Let $g_N$ be the $N$th partial sum. Since 
$(g-g_N)(x)=\langle g-g_N,\mathnormal{\Phi}(\cdot-x)\rangle_\varphi$, application of the Cauchy-Schwarz 
inequality results for any $x\in\mathbb{R}$ in the estimate
 \[
   |g(x)-g_N(x)|\le\|g-g_N\|_\varphi\|\mathnormal{\Phi}(\cdot-x)\|_\varphi
                                     =\|\varphi\|\,\|g-g_N\|_\varphi,
 \]
which proves uniform convergence $g_N\rightarrow g$ on $\mathbb{R}$ and concludes the proof of 
Theorem \ref{Theo_perfrec}.

We note that because of $\mathnormal{\Phi}_\mathrm{int}(n\lambda)=\langle \varphi_\mathrm{dual},
\varphi(\cdot-n\lambda)\rangle$ the interpolating function $\mathnormal{\Phi}_\mathrm{int}$ has the 
\textit{interpolation property}
\begin{equation}
    \mathnormal{\Phi}_\mathrm{int}(n\lambda)=\delta_n,\,n\in\mathbb{Z}. \label{IP_property}
\end{equation}
\noindent\textbf{Remark 2} A similar result holds for the space $V_\lambda(\varphi)$: Every 
$f\in V_\lambda(\varphi)$ can be perfectly reconstructed from its sample values
$f(n\lambda),\,n\in\mathbb{Z},$ by the 
series
\[
   f(x)=\sum_{n=-\infty}^\infty f(n\lambda)\varphi_\mathrm{int}(x-n\lambda),
\]
where the interpolating function $\varphi_\mathrm{int}\in V_\lambda(\varphi)$ is given by
\begin{equation}
   \hat{\varphi}_\mathrm{int}(\xi)=\frac{\hat{\varphi}(\xi)}{\frac{\mathnormal{\Lambda}}{\sqrt{2\pi}}
                             \sum_{n\in {\mathbb Z}}\hat{\varphi}(\xi+n\mathnormal{\Lambda})},\,
			                    \mathnormal{\Lambda}=\frac{2\pi}{\lambda},\label{Rem3}
\end{equation}
provided that the denominator in (\ref{Rem3}) does not vanish (and some additional regularity
assumptions on $\varphi$). This result is due to Walter \cite{Walter} who proved it for orthonormal 
bases $\{\varphi(\cdot-n);n\in\mathbb{Z}\}$. The argument is essentially the same for Riesz bases, 
see \cite{Hammerich2} for the case of a Gaussian function $\varphi$. We shall return to this topic 
at the end of Section 6.

\section{Error Estimate}

If $g$ is in $\mathcal{P}_\varphi\setminus \mathcal{R}_\lambda(\varphi)$, then the series in 
Theorem \ref{Theo_perfrec} still may be computed but will result in an orthogonal 
projection $\tilde{g}$ of $g$ onto $\mathcal{R}_\lambda(\varphi)$ rather than in $g$ itself. The purpose 
of the present section is to estimate the error $|g(x)-\tilde{g}(x)|$ for any $x\in\mathbb{R}$.

Let $g=\boldsymbol{P}_\varphi f$ with $f\in L^2(\mathbb{R})$ and define $\tilde{f}=P_\lambda f$ where 
$P_\lambda$ is the orthogonal projection from $L^2(\mathbb{R})$ onto $V_\lambda(\varphi)$. Then 
$\tilde{g}=\boldsymbol{P}_\varphi \tilde{f}=\boldsymbol{P}_\varphi P_\lambda f$. So, we need to
estimate $|(\boldsymbol{P}_\varphi f)(x)-(\boldsymbol{P}_\varphi P_\lambda f)(x)|$ for arbitrary 
$f\in L^2(\mathbb{R})$ and $x\in\mathbb{R}$.
\begin{proposition} \label{Propo_proj1} Let $\lambda\in\mathbf{\Lambda}$. The orthogonal projection 
$P_\lambda:L^2(\mathbb{R})\rightarrow V_\lambda(\varphi)$ is given by
\begin{equation}
   \widehat{(P_\lambda f)}(\xi)=\left( \mathnormal{\Lambda}
                  \sum_{n\in\mathbb{Z}}\hat{f}(\xi+n\mathnormal{\Lambda})\overline{\hat{\varphi}
                  (\xi+n\mathnormal{\Lambda})}\right) \hat{\varphi}_\mathrm{dual}(\xi),   \label{proj1}
\end{equation}
where $\mathnormal{\Lambda}=2\pi/\lambda$ and $\hat{\varphi}_\mathrm{dual}$ is as in (\ref{dualf2}).
\end{proposition}
\textit{Proof.} In the special case $\lambda=1$ representation (\ref{proj1}) is readily obtained from 
\cite[Theorem 2.9]{Boor}. The general case $\lambda\in\mathbf{\Lambda}$ can then be deduced from the
previous one by rescaling. This concludes the proof of Proposition \ref{Propo_proj1}.

\begin{proposition} \label{Propo_proj2} Let $\lambda\in\mathbf{\Lambda}$. The operator 
$\boldsymbol{Q}_\lambda=\boldsymbol{P}_\varphi P_\lambda: 
L^2(\mathbb{R})\rightarrow \mathcal{R}_\lambda(\varphi)$ is given by
\begin{equation}
   (\boldsymbol{Q}_\lambda f)(x)=\int^\infty_{-\infty}Q_\lambda(x,\xi)\hat{f}(\xi)
                           \overline{\hat{\varphi}(\xi)}\,d\xi,\,x\in\mathbb{R},       \label{proj2_1}
\end{equation}
where
\[
   Q_\lambda(x,\xi)=\frac{\sum_{n\in\mathbb{Z}}e^{ix(\xi+n\mathnormal{\Lambda})}|\hat{\varphi}(\xi+n\mathnormal{\Lambda})|^2}
                                     {\sum_{n\in\mathbb{Z}}|\hat{\varphi}(\xi+n\mathnormal{\Lambda})|^2}.
\]
\end{proposition}
\textit{Proof.} Let $f\in\mathcal{S}({\mathbb R})$ where $\mathcal{S}({\mathbb R})$ is the Schwartz 
space of $C^\infty$ functions on $\mathbb{R}$ rapidly decaying at infinity. By (\ref{OpA}), 
(\ref{dualf2}) and (\ref{proj1}) we get
\begin{eqnarray*}
   (\boldsymbol{P}_\varphi P_\lambda f)(x)&=&\int_{-\infty}^\infty e^{ix\xi}\,\overline{\hat{\varphi}(\xi)}
                                                            \widehat{(P_\lambda f)}(\xi)\,d\xi\\
   &=&\int_{-\infty}^\infty e^{ix\xi}\,\overline{\hat{\varphi}(\xi)}\left\{\left(\mathnormal{\Lambda}\sum_{n\in\mathbb{Z}}
                                                            \hat{f}(\xi+n\mathnormal{\Lambda})
                          \overline{\hat{\varphi}(\xi+n\mathnormal{\Lambda})}\right)\hat{\varphi}_\mathrm{dual}(\xi)\right\}\,d\xi\\
   &=&\mathnormal{\Lambda}\sum_{n\in\mathbb{Z}}\int^\infty_{-\infty}e^{ix(\xi-n\mathnormal{\Lambda})}\,\overline{\hat{\varphi}(\xi)}
                                                                                        \hat{f}(\xi)
                                       \frac{|\hat{\varphi}(\xi-n\mathnormal{\Lambda})|^2}{\mathnormal{\Lambda}\sum_{k\in{\mathbf Z}}
                                                   |\hat{\varphi}(\xi+k\mathnormal{\Lambda})|^2}\,d\xi\\
   &=&\int^\infty_{-\infty}Q_\lambda(x,\xi)\hat{f}(\xi)\overline{\hat{\varphi}(\xi)}\,d\xi.	 
\end{eqnarray*}

Since $P_\lambda$ and $\boldsymbol{P}_\varphi$ are continuous operators from $L^2({\mathbb R})$ into 
itself, so is $\boldsymbol{Q}_\lambda$.  Consequently, the representation (\ref{proj2_1}) extends to 
all of $L^2({\mathbb R})$ by continuity. This completes the proof of Proposition 
\ref{Propo_proj2}.
\begin{lemma}[Generalized Chebyshev Inequality] \label{Lemma_Tscheb} Let $\phi$ be a probability density 
function on $\mathbb{R}$, i.e., $\phi(x)\ge 0$, $\int_\mathbb{R}\phi(x)\,dx$ exists and is equal to $1$. 
Suppose that $M=\int_\mathbb{R}w(|x|)\phi(x)\,dx<\infty$, where $w(x)$ for $x>0$ is positive and 
monotonically increasing. Then
\begin{equation}
   \int_{|x|\ge t}\phi(x)\,dx\le\frac{M}{w(t)}\:\:\:\forall t>0. \label{Tscheb}
\end{equation}
\end{lemma}
\textit{Proof.} Let $t>0$. Then
\begin{eqnarray*}
   \int_{|x|\ge t}\phi(x)\,dx&=&\int_{|x|\ge t}\frac{w(|x|)\phi(x)}{w(|x|)}\,dx\\
                         &\le&\frac{1}{w(t)}\int_{|x|\ge t}w(|x|)\phi(x)\,dx\\
			 &\le&\frac{1}{w(t)}\int_{\mathbb{R}}w(|x|)\phi(x)\,dx\\
			 &=&\frac{M}{w(t)},
\end{eqnarray*}
which proves Lemma \ref{Lemma_Tscheb}. We found the preceding theorem as an exercise in the textbook 
\cite{Feller}. Note that the normalization condition $\int_\mathbb{R}\phi(x)\,dx=1$ is unnecessary.
\begin{theorem} \label{error_est} Let $f\in L^2(\mathbb{R}),\,g=\boldsymbol{P}_\varphi f$ and 
$\tilde{g}=\boldsymbol{Q}_\lambda f$, where $\lambda\in\mathbf{\Lambda}$. Then for all 
$x\in\mathbb{R}$ it holds that
\begin{equation}
   |g(x)-\tilde{g}(x)|^2\le
         8M_w(\varphi)\left(\sum_{n=1}^\infty\frac{1}{w((2n-1)\pi/\lambda)}\right)\|g\|_\varphi^2.
	                                                                              \label{estimate}
\end{equation}
\end{theorem}
\textit{Proof.} By (\ref{OpA}) and (\ref{proj2_1}) we have
\[  
   g(x)-\tilde{g}(x)=
             \int_{-\infty}^\infty(e^{ix\xi}-Q_\lambda(x,\xi))\hat{f}(\xi)\overline{\hat{\varphi}(\xi)}\,d\xi.
\]
Since $|e^{ix\xi}-Q_\lambda(x,\xi)|\le 2E_\lambda(\xi)$, where
\[
   E_\lambda(\xi)=\frac{\sum_{n\in\mathbb{Z}\setminus \{0\}}|\hat{\varphi}(\xi+n\mathnormal{\Lambda})|^2}
                                     {\sum_{n\in\mathbb{Z}}|\hat{\varphi}(\xi+n\mathnormal{\Lambda})|^2},
\]
we obtain by means of the Cauchy-Schwarz inequality that
\[  
   |g(x)-\tilde{g}(x)|^2\le 4\left(\int_{-\infty}^\infty E_\lambda(\xi)|\hat{f}(\xi)|^2\,d\xi\right)
                           \left(\int_{-\infty}^\infty E_\lambda(\xi)|\hat{\varphi}(\xi)|^2\,d\xi\right).
\]
Because
\begin{equation}
   0\le E_\lambda(\xi)\le 1 \label{ineqElambda}
\end{equation}
we get for the first integral factor $\int_\mathbb{R} E_\lambda(\xi)|\hat{f}(\xi)|^2\,d\xi\le \|f\|^2$. 
Writing
\[
   \int_{-\infty}^\infty E_\lambda(\xi)|\hat{\varphi}(\xi)|^2\,d\xi=I(0,\pi/\lambda)+I(\pi/\lambda,\infty),
\]
where we have used the notation
\[
   I(a,b)=\int_{a\le|\xi|<b} E_\lambda(\xi)|\hat{\varphi}(\xi)|^2\,d\xi,
\]
we need to estimate the two constituent integrals. This will be done by means of inequalities 
(\ref{Tscheb}) and (\ref{ineqElambda}).

\noindent\textbf{Integral} $\boldsymbol{I(\pi/\lambda,\infty)}$: 
\begin{equation}
  I(\pi/\lambda,\infty)\le \int_{|\xi|\ge\pi/\lambda} |\hat{\varphi}(\xi)|^2\,d\xi
                                              \le \frac{M_w(\varphi)}{w(\pi/\lambda)}. \label{int1}
\end{equation}

\noindent\textbf{Integral} $\boldsymbol{I(0,\pi/\lambda)}$:
\begin{eqnarray}
   I(0,\pi/\lambda)&=&\int_{-\pi/\lambda}^{\pi/\lambda}(1-E_\lambda(\xi))    
       \left(\sum_{n\in\mathbb{Z}\setminus\{0\}}|\hat{\varphi}(\xi+n\mathnormal{\Lambda})|^2\right)\,d\xi\nonumber\\
       &\le&\sum_{n=1}^\infty\left\{\int_{-\pi/\lambda}^{\pi/\lambda}|\hat{\varphi}(\xi+n\mathnormal{\Lambda})|^2\,d\xi
             +\int_{-\pi/\lambda}^{\pi/\lambda}|\hat{\varphi}(\xi-n\mathnormal{\Lambda})|^2\,d\xi\right\}\nonumber\\
       &\le&\sum_{n=1}^\infty\left\{\int_{-\pi/\lambda}^{\infty}|\hat{\varphi}(\xi+n\mathnormal{\Lambda})|^2\,d\xi
             +\int_{-\infty}^{\pi/\lambda}|\hat{\varphi}(\xi-n\mathnormal{\Lambda})|^2\,d\xi\right\}\nonumber\\
       &=&\sum_{n=1}^\infty\left\{\int_{(2n-1)\pi/\lambda}^{\infty}|\hat{\varphi}(\xi)|^2\,d\xi
             +\int_{-\infty}^{(-2n+1)\pi/\lambda}|\hat{\varphi}(\xi)|^2\,d\xi\right\}\nonumber\\
       &=&\sum_{n=1}^\infty\int_{|\xi|\ge(2n-1)\pi/\lambda}|\hat{\varphi}(\xi)|^2\,d\xi\nonumber\\
       &\le&\sum_{n=1}^\infty \frac{M_w(\varphi)}{w((2n-1)\pi/\lambda)}.\label{int2}
\end{eqnarray}
Because of growth condition (\ref{growth}) imposed on the weight function $w$, the infinite series 
converges. By (\ref{int1}), (\ref{int2}) it follows that
\[
   \int_{-\infty}^\infty E_\lambda(\xi)|\hat{\varphi}(\xi)|^2\,d\xi\le
                2M_w(\varphi)\sum_{n=1}^\infty \frac{1}{w((2n-1)\pi/\lambda)}.
\]
Consequently,
\[
   |g(x)-\tilde{g}(x)|^2\le
         8M_w(\varphi)\left(\sum_{n=1}^\infty\frac{1}{w((2n-1)\pi/\lambda)}\right)\|f\|^2.
\]
By Proposition \ref{Propo_RKHS} we may assume that 
$f\in L^2({\mathbb R}|\,\mathrm{supp}\, \hat{\varphi})$. Then $\|f\|=\|g\|_\varphi$, which concludes the 
proof of Theorem \ref{error_est}.

\subsection{Example: Monomial Weight $w(\xi)=|\xi|^s,\,s>1$}
We define
\[
   \mu_s(\varphi)=\left[\frac{M_w(\varphi)}{\|\varphi\|^2}\right]^\frac{1}{s}.
\]
It is not difficult to prove the identity $\sum_{n=1}^\infty(2n-1)^{-s}=\left(1-2^{-s}\right)\zeta(s)$ 
for the Riemann zeta function $\zeta(s)=\sum_{n=1}^\infty n^{-s},\,\Re(s)>0$. Then, Ineq. 
(\ref{estimate}) turns into
\begin{equation}
   |g(x)-\tilde{g}(x)|^2\le
         8\left(1-2^{-s}\right)\zeta(s)\left[\frac{\mu_s(\varphi)\lambda}{\pi}\right]^s
	                                                            \|\varphi\|^2\|g\|_\varphi^2.
		                                                                        \label{zeta}
\end{equation}
We observe that the right-hand side of (\ref{zeta}) decays order of $s$ to $0$ as 
\begin{equation}
   0\leftarrow\lambda<\lambda_0=\frac{\pi}{\mu_s(\varphi)}. \label{nyq_s}
\end{equation}
Here, of course, $\inf\,\mathbf{\Lambda}=0$ is supposed. Note that the upper bound in (\ref{nyq_s}) 
becomes more and more tight as $s\rightarrow\infty$. In the special case $s=2$ we obtain by means of 
the celebrated identity $\zeta(2)=\pi^2/6$ of Euler\footnote{See M. du Sautoy, \textit{The Music of 
Primes. Why an Unsolved Problem in Mathematics Matters}, Fourth Estate, London, 2003, for a historical 
account} the remarkably simple inequality
\begin{equation}
  |g(x)-\tilde{g}(x)|^2\le
         \left(\mu_2(\varphi)\lambda\right)^2\|\varphi\|^2\|g\|_\varphi^2\:.   \label{allg_ineq}
\end{equation}

\section{Sampling Theorem}
The following theorem is a consequence of Theorem \ref{Theo_perfrec}, Theorem \ref{error_est} and the 
interpolation property (\ref{IP_property}).
\begin{theorem}[Generalized Sampling Theorem] \label{GST} Let the prefilter $\boldsymbol{P}_\varphi$ 
and the set $\mathbf{\Lambda}$ of admissable sampling intervals be as assumed in Section 2. For 
arbitrary $\lambda\in\mathbf{\Lambda}$ the interpolating function 
$\mathnormal{\Phi}_\mathrm{int}\in \mathcal{R}_\lambda(\varphi)\subseteq \mathcal{P}_\varphi$ is defined 
by (\ref{Phi_int}). Then for any signal $g$ in the localization space $\mathcal{P}_\varphi$ of 
$\boldsymbol{P}_\varphi$ the series
\begin{equation}
   \tilde{g}(x)=\sum_{n=-\infty}^\infty g(n\lambda)\mathnormal{\Phi}_\mathrm{int}(x-n\lambda) 
                                                                                      \label{reko_tilde}
\end{equation}
converges in the norm of $\mathcal{P}_\varphi$ and uniformly on $\mathbb{R}$ to a function 
$\tilde{g}\in\mathcal{R}_\lambda(\varphi)$ with the property that 
$\tilde{g}(n\lambda)=g(n\lambda)\:\forall n\in\mathbb{Z}$. For all other $x\in\mathbb{R}$ the relative 
approximation error $\epsilon_\lambda(x)=|g(x)-\tilde{g}(x)|/\|g\|_\varphi$ satisfies the estimate
\begin{equation}
  \epsilon_\lambda^2(x)\le 8M_w(\varphi)\sum_{n=1}^\infty\frac{1}{w((2n-1)\pi/\lambda)}.\label{est_a}
\end{equation}
\end{theorem}
Theorem \ref{GST} may serve as pattern for the proof of specific sampling theorems. Examples 
are given in the following subsection.

Special attention deserves the upper bound in Ineq. (\ref{est_a}). As seen in Section 5.1, it may allow 
for the identification of a \textit{critical sampling interval} $\lambda_0$ with the property that the 
error $\epsilon_\lambda(x)$ diminishes substantially as soon as the sampling interval $\lambda$ falls 
under $\lambda_0$. Similarly to the Nyquist interval in the classical sampling theorem \cite{Shannon}, 
\cite{Vaid}, we may also get a link between $\lambda_0$ and some measure of bandwidth for the prefilter 
function $\varphi$ (or the prefilter $\boldsymbol{P}_\varphi$). Indeed, inequality (\ref{allg_ineq}) 
suggests the critical sampling interval $\lambda_0=1/\sigma(\varphi)$, where 
$\sigma(\varphi)=\mu_2(\varphi)$ is defined by
\begin{equation}
  \sigma^2(\varphi)=\|\varphi\|^{-2}\int_{-\infty}^\infty \xi^2|\hat{\varphi}(\xi)|^2\,d\xi.
\end{equation}
$\sigma(\varphi)$ may be viewed as the standard deviation of the probability 
density function $\|\hat{\varphi}\|^{-2}|\hat{\varphi}(\cdot)|^2$ in the frequency domain. Following
Gabor \cite{Gabor} where a centered version of $\sigma(\varphi)$ called "effective frequency width" has 
been defined, we might take $\sigma(\varphi)$ as a measure of "soft bandwidth" for $\varphi$ (or 
$\boldsymbol{P}_\varphi$). Inequality (\ref{allg_ineq}) then tells us that, under the assumption 
$\sigma(\varphi)<\infty$, \textit{the sampling interval should always be chosen smaller than the 
reciprocal "soft bandwidth" $\sigma(\varphi)$}.

A definition of "soft bandwidth" could, of course, also be based on other weights $w$ leading to, e.g., 
the moment-like quantities $\mu_s(\varphi)$ of Section 5.1.

\subsection{Instances of the Generalized Sampling Theorem}
\noindent 1) The prefilter function
\begin{equation}
	\varphi(x)=\mbox{sinc}\,(\pi\beta x)	
		\:\:\stackrel{\mathrm{Fourier}}{\longleftrightarrow}\:\:
		\hat{\varphi}(\xi)=\frac{1}{\sqrt{2\pi}\beta}\chi_{[-\pi\beta,\pi\beta]}(\xi),
		                                                                    \label{ideal}	
\end{equation}
where $\mbox{sinc}\,x=(\sin x)/x$, defines the ideal low-pass filter of (two-sided) bandwidth $2\pi\beta$. For an arbitrary weight function
$w_s(\xi)=s(\pi\beta)^{1-s}|\xi|^{s-1}$ with $s>2$ we compute that $M_{w_s}(\varphi)=1/\beta$. The Riesz 
basis condition (\ref{Riesz}) is fulfilled for $\lambda\in \mathbf{\Lambda}=[1/\beta,\infty)$. 
Localization space $\mathcal{P}_\varphi$ coincides with the space of finite-energy signals 
bandlimited to $[-\pi\beta,\pi\beta]$. With the special choice $\lambda=1/\beta$ (Nyquist interval) we now 
obtain by means of Theorem \ref{GST} for fixed $g\in\mathcal{P}_\varphi$ and an arbitrary weight 
function $w_s(\xi),\,s>2,$ observing that $\|g\|_\varphi=\beta\|g\|$, the inequality 
\[
   |g(x)-\tilde{g}(x)|^2\le\frac{8\beta}{s}(1-2^{-s+1})\zeta(s-1)\|g\|^2.
\]
Since the upper bound tends to 0 as $s\rightarrow \infty$, it follows that  
$\tilde{g}(x)=g(x)\,\forall x\in\mathbb{R}$. Furthermore, by (\ref{Phi_int}) we immediately see that  
$\mathnormal{\Phi}_\mathrm{int}(x)=\mbox{sinc}\,(\pi\beta x)$. As a result, we have proved the 
classical sampling theorem \cite{Shannon} showing that it is within the scope of our generalized 
approach.  
  
\noindent 2) For the prefilter function 
\begin{equation}
	\varphi(x)=\frac{1}{\sqrt{2\pi}(1/\beta)}e^{-\frac{x^2}{2(1/\beta)^2}}	
		\:\:\stackrel{\mathrm{Fourier}}{\longleftrightarrow}\:\:
		\hat{\varphi}(\xi)=\frac{1}{\sqrt{2\pi}}e^{-\frac{\xi^2}{2\beta^2}},	\label{Gaussian}	
\end{equation}
a Gaussian probability density function of standard deviation $1/\beta,\,\beta>0,$ the weight 
function
\[
   w(\xi)=e^\frac{\xi^2}{2\beta^2}
\]
and $\mathbf{\Lambda}=\mathbb{R}_+$ may be used. Since $M_w(\varphi)=\beta/\sqrt{2\pi}<\infty$ and for 
$\lambda>0$ small enough (condition (\ref{NyqGauss}), see below, is sufficient)
\[
  \sum_{n=1}^\infty\frac{1}{w((2n-1)\pi/\lambda)}=\sum_{n=1}^\infty 
                 e^{-\frac{(\pi/\lambda)^2}{2\left(\frac{\beta}{2n-1}\right)^2}}
                                            <2e^{-\frac{(\pi/\lambda)^2}{2\beta^2}},
\] 
we obtain by Theorem \ref{GST} the estimate
\begin{equation}
   \left[|g(x)-\tilde{g}(x)|/\|g\|_\varphi\right]^2\le \frac{16\beta}{\sqrt{2\pi}}
                              e^{-\frac{(\pi/\lambda)^2}{2\beta^2}}
                                        \:\:\:\forall x\in \mathbb{R}.  \label{Gauss_error}
\end{equation}
Because of the "three-sigma rule" for the Gaussian function in the interpretation given in 
\cite{Hammerich1}, the right-hand side of Ineq. (\ref{Gauss_error}) becomes small as soon as 
$\pi/\lambda>\pi\beta$  or, equivalently, 
\begin{equation}
   0<\lambda<\lambda_0=\frac{1}{\beta},  \label{NyqGauss}
\end{equation}
then decaying super-exponentially to 0 as $\lambda\rightarrow 0$. (By means of a specific inequality for
the tails of the Gaussian function it is shown in \cite{Hammerich1} that the error
becomes very small already when $\lambda=\lambda_0=1/\beta$.) We note that $\lambda_0$ is related to 
the "soft bandwidth" $\sigma(\varphi)$ of the Gaussian function (\ref{Gaussian}) by the equation 
$\lambda_0=1/(\sqrt{2}\sigma(\varphi))$. The corresponding 
interpolating function $\mathnormal{\Phi}_\mathrm{int}$ has been calculated in \cite{Hammerich1}, 
\cite{Hammerich2}. 

\noindent3) The centered B-spline of order $m$ \cite{Unser99}, $m=2,3,\ldots,$
\begin{eqnarray}
   &\beta^{m-1}(x)&=(\underbrace{\beta^0*\ldots*\beta^0}_{m\,\mbox{times}})(x),\,
	       \beta^0(x)=\chi_{\left[-\frac{1}{2},\frac{1}{2}\right]}(x)  \label{cspline}\\	
   & & \stackrel{\mathrm{Fourier}}{\longleftrightarrow}\hat{\beta}^{m-1}(\xi)
                         =\frac{1}{\sqrt{2\pi}}\left(\frac{\sin(\xi/2)}{\xi/2}\right)^m,\nonumber	
\end{eqnarray}
may also be used as prefilter function $\varphi$. Any weight function
\[
   w(\xi)=|\xi|^s\,,1<s<2m-1,
\]
may be taken. Since the central term in (\ref{Riesz}) has zeros when 
$\lambda=\frac{1}{2},\frac{1}{3},\frac{1}{4},\ldots$, we choose 
$\mathbf{\Lambda}=\mathbb{R}_+\setminus\{\frac{1}{2},\frac{1}{3},\frac{1}{4},\ldots\}$ as the set of
admissable sampling intervals. Then Theorem \ref{GST} applies and Ineq. (\ref{zeta}) becomes 
\begin{equation}
   |g(x)-\tilde{g}(x)|^2\le C\,\lambda^s\:\:\:\forall x\in\mathbb{R},    \label{err_spline}
\end{equation}
where $C$ is some finite constant not depending on $\lambda\in\mathbf{\Lambda}$. Now, let 
$\mathbf{\Lambda}\ni\lambda\rightarrow \lambda_\ell=1/\ell$ for $\ell\in\{2,3,\ldots\}$ held constant. 
Since zeros arising in (\ref{Phi_int}) in the denominator are cancelled by corresponding 
zeros of the numerator $\hat{\mathnormal{\Phi}}(\xi)=\hat{\beta}^{2m-1}(\xi)$ we
observe convergence $\mathnormal{\Phi}_\mathrm{int}\rightarrow \mathnormal{\Phi}_{\ell,\mathrm{int}}$ in 
$L^2(\mathbb{R})$ where $\mathnormal{\Phi}_{\ell,\mathrm{int}}$ is given by
\[
   \hat{\mathnormal{\Phi}}_{\ell,\mathrm{int}}(\xi)=\frac{\hat{\mathnormal{\Phi}}_\ell(\xi)}
       {\frac{\mathnormal{\Lambda}_\ell}{\sqrt{2\pi}}
               \sum_{n\in {\mathbb Z}}\hat{\mathnormal{\Phi}}_\ell(\xi+n\mathnormal{\Lambda}_\ell)},\,
	                                            \mathnormal{\Lambda}_\ell=\frac{2\pi}{\lambda_\ell},
\]
with $\hat{\mathnormal{\Phi}}_\ell(\xi)=(1/\ell)\hat{\mathnormal{\Phi}}(\xi/\ell)$. In Figure 1(a)-(d) 
this bevaviour is depicted for prefilter function $\varphi(x)=\beta^2(x)$ in case $\ell=4$.

\setlength{\unitlength}{1cm}
\vspace{1cm}
\begin{minipage}[t]{5.0cm}
\epsfig{file=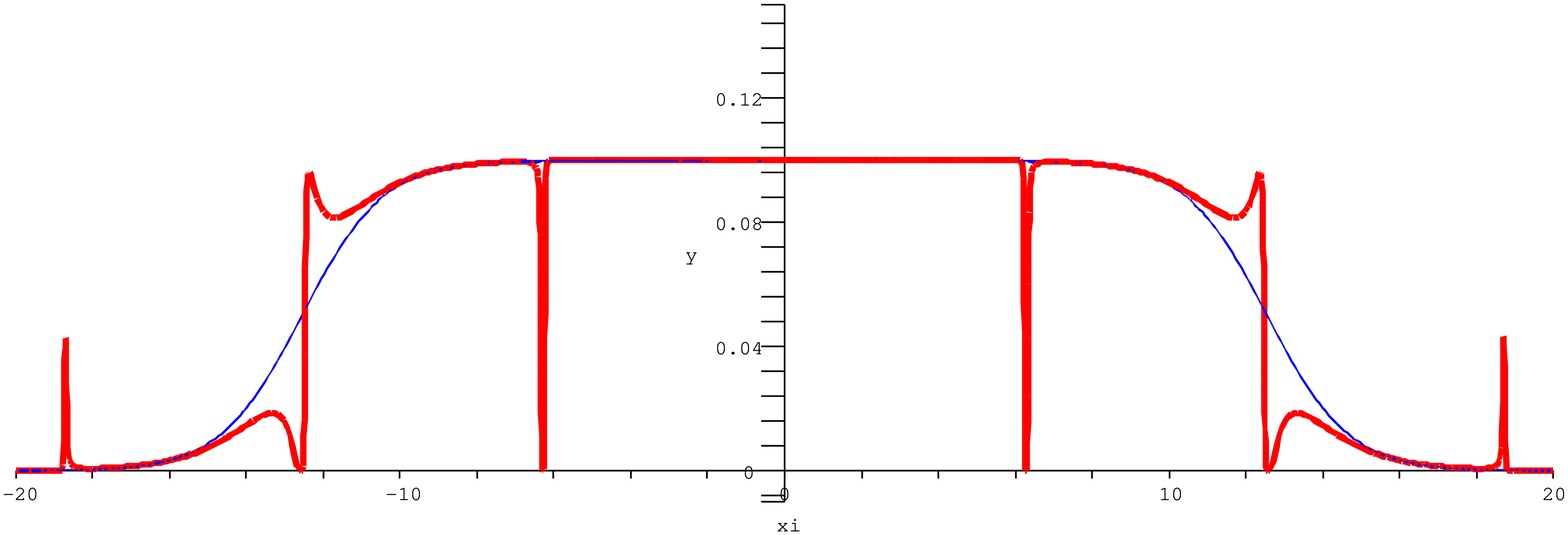, width=\textwidth} \par
\vspace{-0.5cm}
\begin{center} (a) \end{center}
%\caption{Interpolating function $\Phi_\mathrm{int}$ (order $m=3,\lambda=1/4$)}
\end{minipage}\hfill
\begin{minipage}[t]{5.0cm}
\epsfig{file=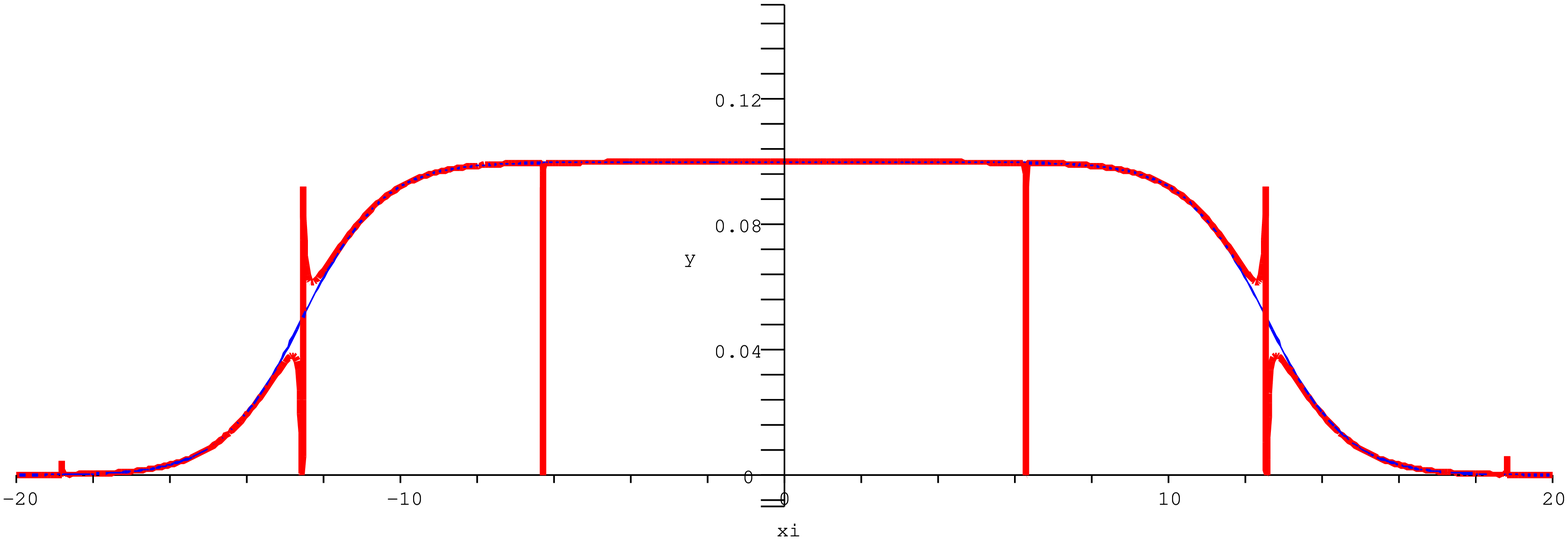, width=\textwidth} \par
\vspace{-0.5cm}
\begin{center} (b) \end{center}
%\caption{(b) Interpolating function $\Phi_\mathrm{int}$ (order $m=3,\lambda=1/4$)}
\end{minipage}

\vspace{0.5cm}

\setlength{\unitlength}{1cm}
\begin{minipage}[t]{5.0cm}
\epsfig{file=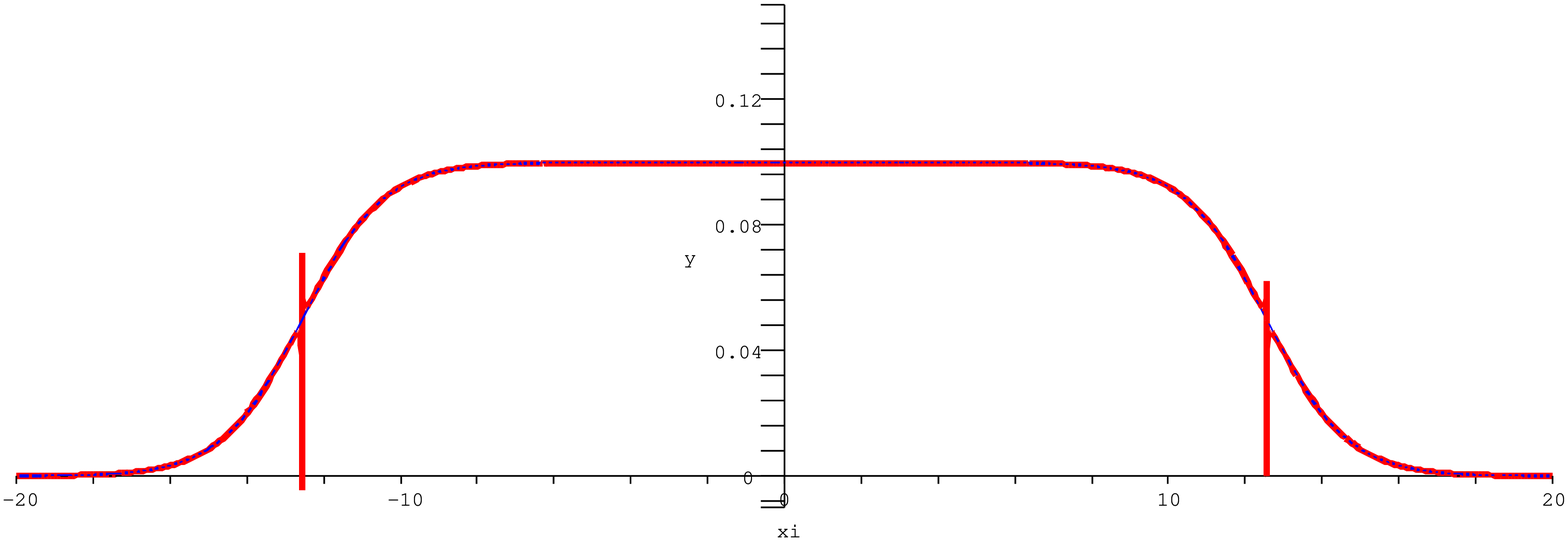, width=\textwidth} \par
\vspace{-0.5cm}
\begin{center} (c) \end{center}
%\caption{(c) Interpolating function $\Phi_\mathrm{int}$ (order $m=3,\lambda=1/4$)}
\end{minipage}\hfill
\begin{minipage}[t]{5.0cm}
\epsfig{file=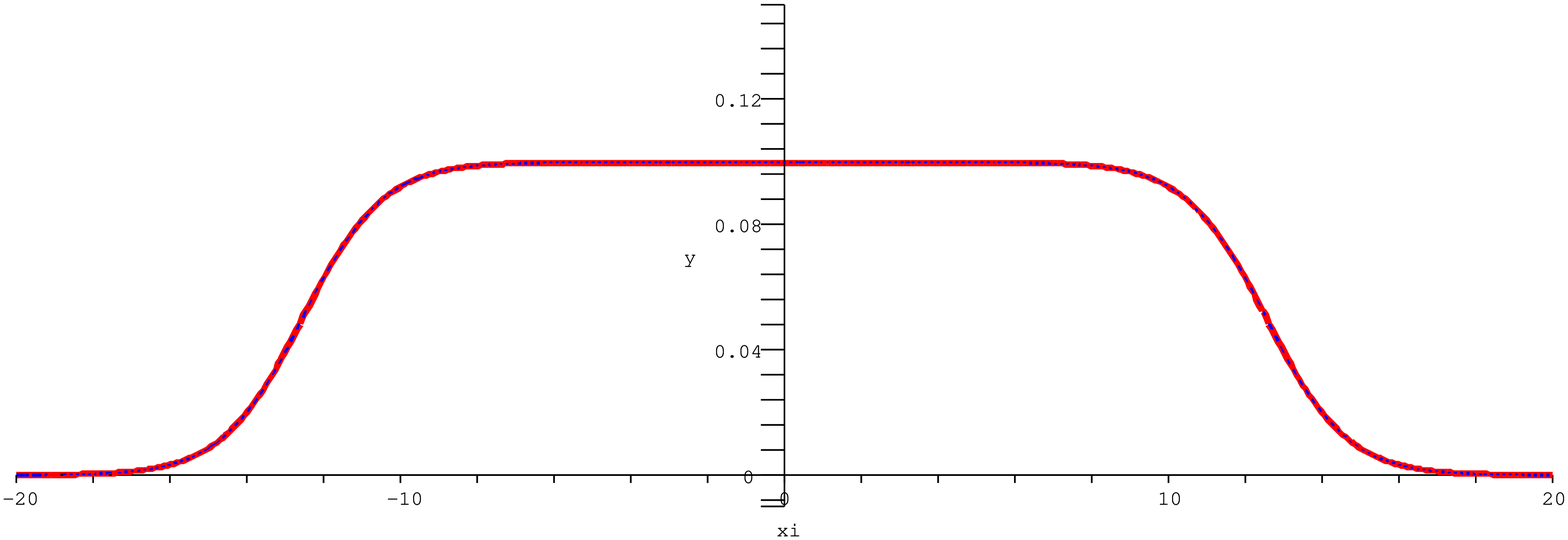, width=\textwidth} \par
\vspace{-0.5cm}
\begin{center} (d) \end{center}
%\caption{Interpolating functions: Gaussian, B-spline, Ideal}
\end{minipage}
\nopagebreak[4]\begin{center} {\bf Fig. 1.} $\mathnormal{\Phi}_{\ell,\mathrm{int}}\,(\ell=4)$ and  
  $\mathnormal{\Phi}_\mathrm{int}$ as $\lambda\rightarrow 1/4$. \end{center}
\vspace{1cm}

\noindent As a consequence, $\tilde{g}$ as given by (\ref{reko_tilde}) tends to $\tilde{g}_\ell$ in 
$L^2(\mathbb{R})$ as $\lambda\rightarrow \lambda_\ell$, where
\[
   \tilde{g}_\ell(x)=\sum_{n\in{\mathbb Z}}g(n/\ell)\mathnormal{\Phi}_{\ell,\mathrm{int}}(x-n/\ell).
\]
Since inequality (\ref{err_spline}) continues to hold as 
$\lambda\rightarrow\lambda_\ell,\,\tilde{g}\rightarrow\tilde{g}_\ell$, it follows that 
\begin{equation}
   |g(x)-\tilde{g}_\ell(x)|^2\le C\,\ell^{-s}\:\:\:\mbox{a.e.}  \label{ae}
\end{equation}
for any $\ell\in\mathbb{N}$ held constant. Because $g,\,\tilde{g}$ are continuous functions, 
inequality (\ref{ae}) holds everywhere in $\mathbb{R}$. By inspection of the Fourier transform of 
$\tilde{g}_\ell$ it can be seen that always $\tilde{g}_\ell\in \mathcal{P}_\varphi$ (whereas 
$\mathnormal{\Phi}_{\ell,\mathrm{int}}\notin \mathcal{P}_\varphi,\,\ell=2,3,\ldots$!). As a result, the 
functions $\tilde{g}_\ell$ form an approximation to $g$ in $\mathcal{P}_\varphi$ with $s$th-order decay 
of squared error $|g(x)-\tilde{g}_\ell(x)|^2$ as $\ell\rightarrow\infty$, uniformly in 
$x\in\mathbb{R}$.

Since $\mathnormal{\Phi}_{\ell,\mathrm{int}}$ is the interpolating function (\ref{Phi_int}) for the 
dilated prefilter function $\varphi_\ell(x)=\ell^{1/2}\beta^{m-1}(\ell x)$ and $\lambda=1/\ell$, it can 
be computed by the theorem of residues as indicated in \cite{AldroubiUnserEden}.

There is no reason to refrain from using non-centered B-spline functions $N_m(x)$ \cite{Schoenberg} of 
any order $m=2,3,4,\ldots$ defined by
\begin{eqnarray}
   &N_m(x)&=(\underbrace{\chi_{[0,1]}*\ldots*\chi_{[0,1]}}_{m\,\mbox{times}})(x)=
                                                              \beta^{m-1}\left(x-\frac{m}{2}\right) 
                                                                                      \label{ncspline}\\	
& & \stackrel{\mathrm{Fourier}}{\longleftrightarrow}\hat{N}_m(\xi)
             =\frac{1}{\sqrt{2\pi}}\left(e^{-i\xi/2}\frac{\sin(\xi/2)}{\xi/2}\right)^m.      \nonumber
\end{eqnarray}
Indeed, the prefilter function $\varphi(x)=N_m(x)$ will result in the same localization space 
$\mathcal{P}_\varphi$ as when using $\varphi(x)=\beta^{m-1}(x)$. The interpolating functions 
$\mathnormal{\Phi}_\mathrm{int}$ will also coincide. Therefore, when $\varphi(x)=N_m(x)$ we 
always have a sampling theorem in the space 
$\mathcal{R}_\lambda(\varphi)=\boldsymbol{P}_\varphi V_\lambda(\varphi)$ (at least when 
$0<\lambda\not=\frac{1}{2},\frac{1}{3},\frac{1}{4},\ldots$). On the other hand, there is no sampling 
theorem in the shift-invariant subspace $V_\lambda(\varphi)\subseteq L^2(\mathbb{R})$ in case 
$\lambda=1$ and $m=3,5,7,\ldots$ as shown by the following counterexample (a simple generalization of 
the one given by Walter \cite{Walter}): The interpolating function (see Remark 2) 
$\varphi_\mathrm{int}\in V_1(N_m)$, $m\ge3$ odd, defined by
\[
   \hat{\varphi}_\mathrm{int}(\xi)=\frac{\hat{N}_m(\xi)}
          {\sqrt{2\pi}\sum_{n\in {\mathbb Z}}\hat{N}_m(\xi+2\pi n)}
\]
has a pole in $\xi=\pi$ because in the numerator we obtain $\hat{N}_m(\pi)\not=0$ whereas the 
denominator becomes
\[
   \sqrt{2\pi}\sum_{n=-\infty}^\infty\hat{N}_m(\pi+2\pi n)
	   =\sum_{n=-\infty}^\infty\left[i\left(n+\frac{1}{2}\right)\pi\right]^{-m}=0.
\]

As a conclusion, in Figure 2 the Fourier transforms of related interpolating functions 
$\mathnormal{\Phi}_\mathrm{int}$ or $\mathnormal{\Phi}_{\ell,\mathrm{int}}$ are depicted: for 
ideal low-pass ($\beta=4$), Gaussian ($\beta=2$) and B-spline ($m=3$) prefilter function $\varphi$ and 
$\lambda=0.25$ or $\ell=4$.

\setlength{\unitlength}{1cm}
\vspace{1cm}
\hspace{3cm}
\begin{minipage}[t]{5.0cm}
\epsfig{file=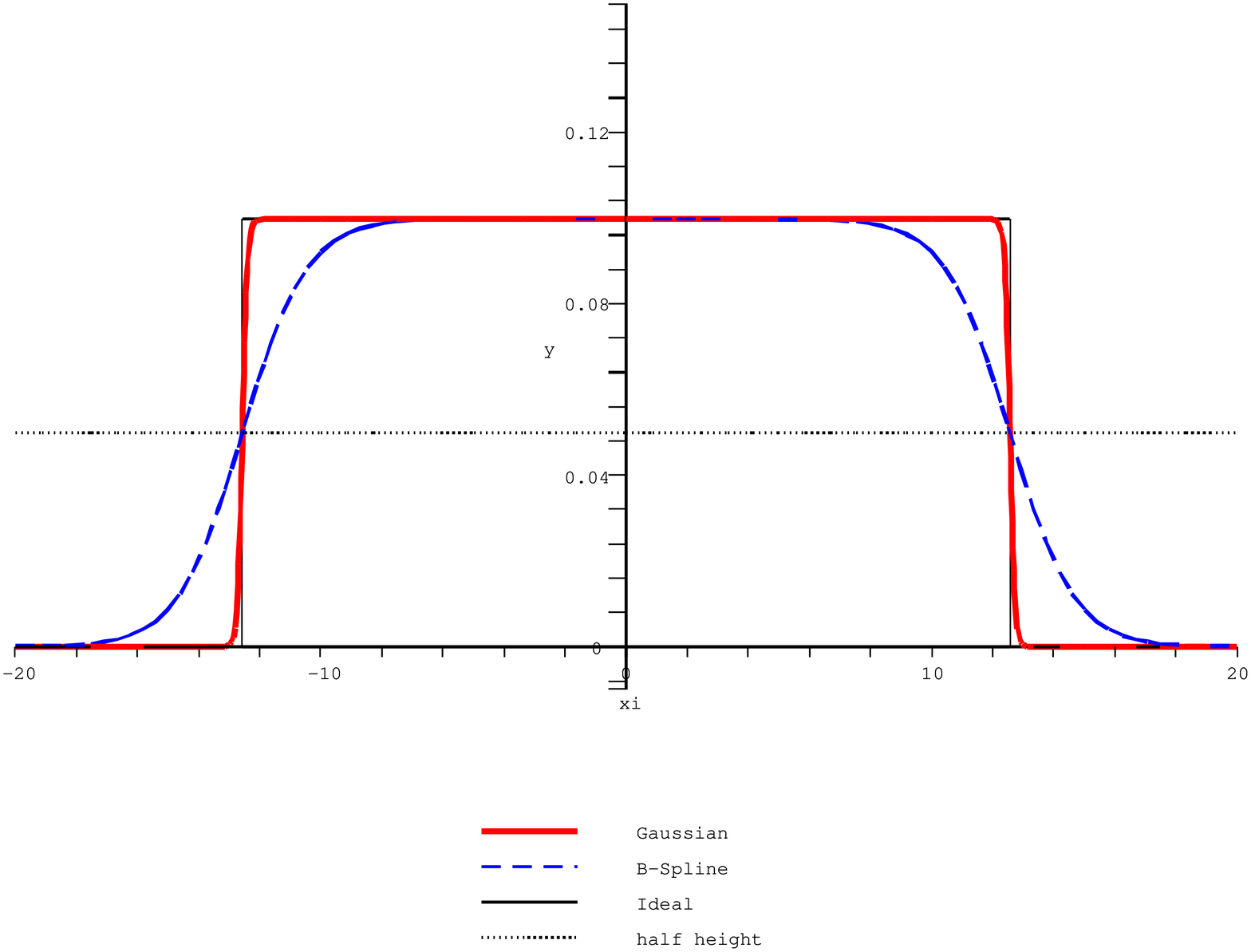, width=\textwidth} \par
\end{minipage}
\nopagebreak\begin{center}{\bf Fig. 2.} Interpolating functions in frequency domain.\end{center}
\vspace{1cm}

\begin{center} \textsc{ACKNOWLEDGEMENT}\end{center}

\noindent The present work originates in a talk given in the Time-Frequency Seminar of EUCETIFA during 
       a stay at the University of Vienna, May 15-17, 2006. We are grateful to K. Gr\"ochenig and 
       H. G. Feichtinger for hospitality and advice.

\end{document}